\documentclass[letterpaper]{jpconf}
\usepackage{graphicx}
\begin{document}
\title{Non-photonic electron-hadron correlations in Cu+Cu at $\sqrt{s_{NN}}$ = 200 GeV}

\author{Miroslav Kr\r us, for the STAR Collaboration}

\address{Czech Technical University in Prague, FNSPE, B\v rehov\' a~7, 115 19 Prague, Czech Republic}
\address{Nuclear Physics Institute of the AS CR v. v. i., 250 68 \v Re\v z, Czech Republic}
\begin{abstract}
In this paper we present the study of the azimuthal correlation function of non-photonic electrons with low-p$_\mathrm{T}$ hadrons produced in Cu+Cu collision at $\sqrt{s_{NN}}$~=~200~GeV measured by STAR experiment at RHIC.  Possible modification of the awayside peak is observed.
\end{abstract}

\section{Introduction}
Recent STAR and PHENIX experimental results \cite{horner, adler, energ1, energ2} have shown that light partons with
high transverse momentum lose a significant amount of energy when
traversing the dense nuclear medium created in central Au+Au collisions.
This leads to suppression of high-p$_\mathrm{T}$ hadron yields and the
disappearance of the away-side peak in the azimuthal hadron-hadron
correlation function. The investigation of the azimuthal correlation function of high-p$_\mathrm{T}$ hadrons with medium- or low-p$_\mathrm{T}$ associated particle have shown  a broad double-peak structure on the away side ($\Delta \phi = \pi$)\cite{abelev}. There is an intensive discussion of the possible explanation of this observation including Mach cone scenario \cite{machcone}, gluon Cherenkov radiation \cite{cherenkov}, or jet deflection \cite{deflect}.

An interesting question is whether similar effect is present in a case of heavy quarks passing the medium. Heavy quarks are primarily produced during early stages of a nuclear collision, then they interact with the medium, and they can be used as a probe of the space-time evolution of the medium arising from a heavy ion collision. Because of their large masses, their energy loss is expected to be influenced by dead-cone effect \cite{doks} and elastic collision energy loss \cite{must}.
\section{Electron identification}
\label{identification}
The direct identification of heavy quark hadrons is difficult with current detectors and RHIC luminosities. Therefore most previous studies used non-photonic electrons consisting of electrons from semileptonic decays of charm and bottom hadrons. A strong suppression of non-photonic electrons, similar to inclusive hadrons, was observed in central Au+Au collisions \cite{jaro}. The identification uses semileptonic decays of open heavy flavour hadrons  (e.g. $D^0 \rightarrow e^{+}K^{-}\nu_e$) over broad p$_\mathrm{T}$-range. These non-photonic electrons keep well direction of the mother heavy hadron when electron has p$_\mathrm{T} > $ 3 GeV/c \cite{gang}. The purity of electron sample, for electrons with p$_\mathrm{T}$ in range 3 GeV/c - 6 GeV/c, is above 99\% \cite{gang}. 

For results presented here, the 0-20\% central\cite{central} Cu+Cu collisions at $\sqrt{s_{NN}}$ = 200 GeV data measured in 2005 by STAR are used. The analysis steps of the electron identification has been reported in details in \cite{jaro}. In general, charged particle tracks are reconstructed in the TPC \cite{tpc} (Time Projection Chamber) and electrons can be identified by their TPC ionization energy loss (3.31 keV/cm $<~dE/dx~<$ 4.64 keV/cm). To enlarge the yield of high-p$_\mathrm{T}$ electrons, the high-tower trigger requires, at least, one track's projection into the BEMC \cite{barrel} (Barrel Electromagnetic Calorimeter) with the deposition of its energy above a threshold in a single BEMC tower. For Cu+Cu, the high-tower threshold is 3.75 GeV. For electron identification, the $p/E_{TOW}$ cut was used, where $p$ is the TPC momentum of the track and $E_{TOW}$ is energy deposited in the BEMC tower. This ratio for electrons should be 1. To identify electons, the 0.3 $ <~p/E_{TOW}~<$ 1.5 cut was used. Electron yield can be also enlarged by the use of the SMD \cite{barrel} (Shower Maximum Detector), where electrons are identified by their electromagnetic shower profile that is more broader than for hadrons (cluster size for electrons $\geq$ 2). To avoid the large amount of photon conversion electron background, the pseudorapidity range is 0 $<~\eta~<$ 0.7. 

The major difficulty in the electron analysis is the fact that there are many sources of the electrons other than semileptonic decays of heavy mesons, for instance from photon conversions in the detector material between the interaction point and the TPC and $\pi^0$ and/or $\eta$ Dalitz decays. These photonic electrons are identified by the invariant mass distribution of electron-positron pairs. The photonic electron yield is given by the difference between the opposite- and same-sign distribution below the invariant mass cut (typically $\pi^0$ mass). The same-sign distribution is due to combinations of random pairs.

\section{Non-photonic electron-hadron correlations}
\label{correlations}
The study of the azimuthal non-photonic electron-hadron correlations (trigger electron with $3<\mathrm{p}_{\mathrm{T}}^{\mathrm{trig}}<6$ GeV/c and associated charged hadron with 0.15$< \mathrm{p}_{\mathrm{T}}^{\mathrm{assoc}}<0.5$ GeV/c) uses the semi-inclusive electron sample \cite{gang}, when electrons with the opposite-sign partner with an invariant mass cut are excluded from the inclusive electron sample. The non-photonic electron-hadron correlations ($\Delta \Phi_{NP}$) are obtained by the formula
\begin{displaymath}
 \Delta \Phi_{NP}=\Delta \Phi_{SI}+\Delta \Phi_{SS}-\left( \frac{1}{\varepsilon}-1 \right) \left( \Delta \Phi_{OS}-\Delta \Phi_{SS} \right),
\end{displaymath}
where each term represents correlation functions of the individual electron samples (SI - semi-inclusive, SS - same-sign, and OS - opposite-sign) with charged hadrons and $\varepsilon$ is the efficiency of the photonic electron reconstruction estimated by embedding simulated data into real events ($\varepsilon$ = 0.665). Eventualy the elliptic flow contribution to the correlation function must be subtracted . This contribution is given by
\begin{displaymath}
 \frac{1}{N_{trig}}\frac{dN}{d\Delta \phi}= A \left( 1+2v_2^e v_2^h cos(2\Delta \phi)\right),
\end{displaymath}
 where A is determined by the ZYAM \cite{adler} method, $v_2^e$ is electron and $v_2^h$ hadron  elliptic flows. We assume that for the most central Cu+Cu collisions $v_2^e$ = $v_2^h$ = 0.05. 

\begin{figure}[hbtp]
\begin{center}
 \includegraphics[width = 15 cm]{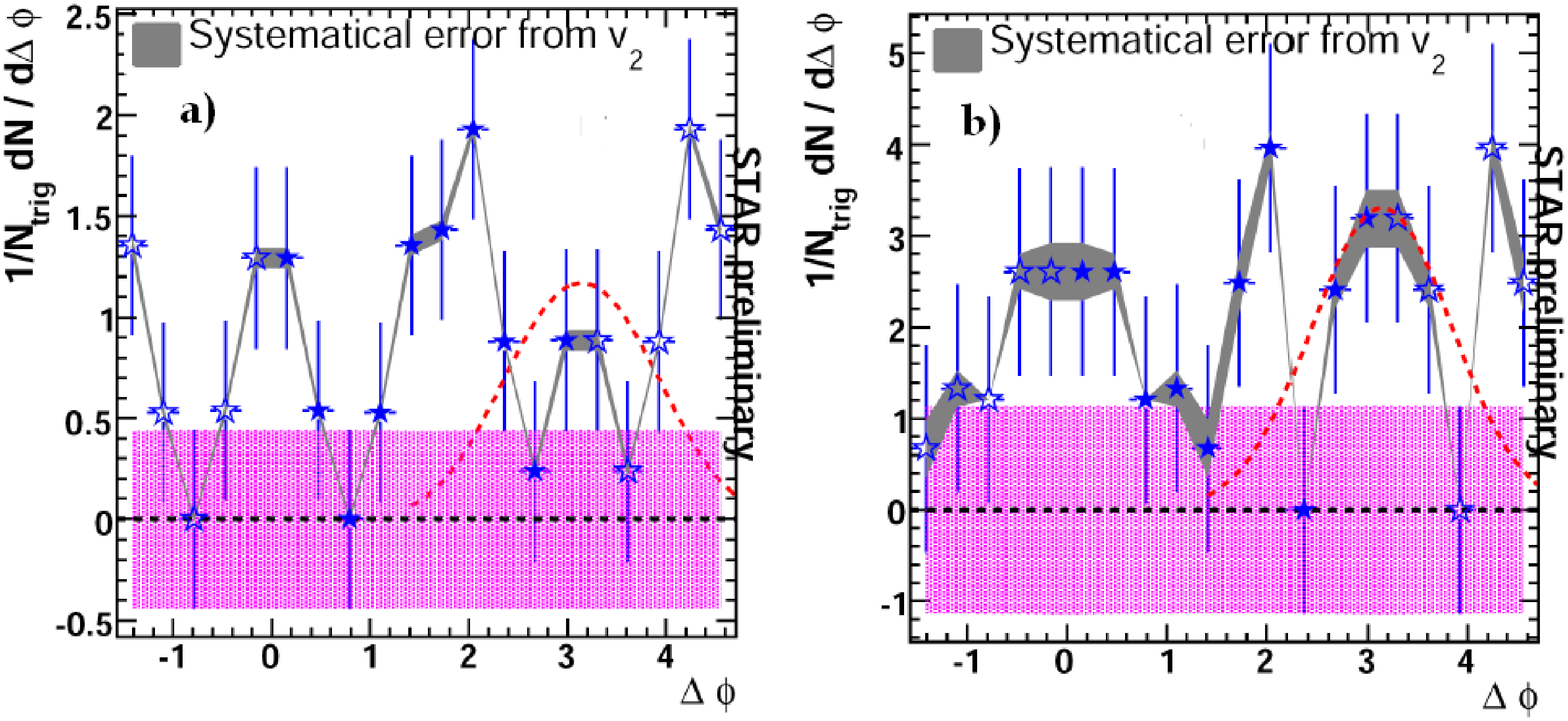}

\caption{Non-photonic electron-hadron correlations. Panel a) shows correlation in Cu+Cu collisions at 200 GeV, panel b) in Au+Au collisions at 200 GeV after $v_2$ subtraction ($v_2$ = 0.05). The error bars are statistical and the error band represents ZYAM systematical uncertainty. The dashed curve is the PYTHIA prediction of the away side peak.}
\label{pict}
\end{center}

\end{figure}

Despite large statistical errors, one can see clear correlation structure of non-photonic electrons with hadrons in central (centrality 0 - 20\%) Cu+Cu collisions (Fig. \ref{pict}). On the nearside ($\Delta \phi$ = 0), the single peak  represents the heavy quark fragmentation and possible interaction with medium. On the awayside ($\Delta \phi$ = $\pi$), the correlation function shows a broad modification or possible double-hump structure.  Similar pattern can been seen in central Au+Au collisions (centrality 0 - 20\%) \cite{gang} (Fig. \ref{pict}).
\newpage
\section{Conclusion}
\label{concl}

The non-photonic azimuthal electron-hadron correlations at $\sqrt{s_{NN}}$ = 200 GeV were measured in the 20\% most central Cu+Cu collisions by STAR. Within large statistical errors, the comparision of the awayside peak with PYTHIA prediction indicates the broad modification of the awayside peak in both Cu+Cu and Au+Au central collisions. This modification is similar to that was observed in the di-hadron correlations in Au+Au. This result may indicate similar response of the nuclear matter produced in central collisions to passage of heavy quarks as well of the light ones.

\section*{Acknowledgments}
\label{acknow}
This work was supported in part by the IRP AV0Z10480505, by GACR grant 202/07/0079 and by grants LC07048 and LA09013
of the Ministry of Education, Youth and Sports of the Czech Republic.

\end{document}